\documentclass[11pt]{wlscirep}
\usepackage{graphicx}
\usepackage{hyperref}

\title{Pair-correlation ansatz for the ground state of interacting bosons in an arbitrary one-dimensional potential}
   
\author[1]{Przemys\l aw Ko\'{s}cik}
\author[2,3]{Arkadiusz Kuro\'{s}}
\author[1]{ Adam Pieprzycki}
\author[4,*]{Tomasz Sowi\'{n}ski}
\affil[1]{University of Applied Sciences, Department of Computer Sciences, ul. Mickiewicza 8, PL-33100 Tarn\'{o}w, Poland}
\affil[2]{Institute of Physics, Jan Kochanowski University, ul. Uniwersytecka 7, PL-25406 Kielce, Poland}
\affil[3]{Instytut Fizyki Teoretycznej, Uniwersytet Jagielloński, ul. Łojasiewicza 11, PL-30348 Kraków, Poland}
\affil[4]{Institute of Physics, Polish Academy of Sciences, Aleja Lotnikow 32/46, PL-02668 Warsaw, Poland}
\affil[*]{tomsow@ifpan.edu.pl}

\begin{abstract}
We derive and describe a very accurate variational scheme for the ground state of the system of a few ultra-cold bosons confined in one-dimensional traps of arbitrary shapes. It is based on assumption that all inter-particle correlations have two-body nature. By construction, the proposed ansatz is exact in the noninteracting limit, exactly encodes boundary conditions forced by contact interactions, and gives full control on accuracy in the limit of infinite repulsions. We show its efficiency in a whole range of intermediate interactions for different external potentials. Our results manifest that for generic non-parabolic potentials mutual correlations forced by interactions cannot be captured by distance-dependent functions. 
\end{abstract}
\begin{document}

\flushbottom
\maketitle
\section*{Introduction}

At the moment the quantum mechanics was rigorously formulated, it became obvious that understanding of different properties of quantum many-body systems is incomparably more challenging than their one- and two-particle counterparts\cite{1971FetterBook}. The main source of this obstacle is of course the tensor product structure of the many-body Hilbert space growing exponentially with the number of particles. That simple fact, already for systems containing only several particles in one spatial dimension, leads directly to an impassable fiasco of any accurate numerical approach willing to take into account all possible many-body configurations coupled by mutual interactions. In fact, even after firm reduction of allowed states (justified only for relatively weak interactions), direct methods like the exact diagonalization \cite{2008WeisseEDChapter} are generally time-consuming and not efficient. Of course, some clever modifications of these approaches may shift the boundary of their applicability \cite{1998HaugsetPRA,2007DeuretzbacherPRA,2018JeszenskiPRA,2018KoscikPhysLettA,2019ChrostowskiAPPA}. Still, however, the main stumbling block, {\it i.e.}, the exponential growth of the Hilbert space, remains unchanged.
Moreover, only few exactly solvable models for many-body systems are known like: the Moshinsky model\cite{1968MoshinskyAJP,1985BialynickiLetMPhys}, the Lieb-Liniger model\cite{1963LiebPR,1963LiebPRb}, the McGuire-Gaudin-Yang model \cite{1966McGuireJMP1,1966McGuireJMP2,1967GaudinPLA,1967YangPRL,2020GamayunSciPost}, the Calogero-Sutherland model\cite{1971CalogeroJMP,1971SutherlandJMP}, the Yang–Baxter integrable models \cite{2016BatchelorJPhysA}, or the one-dimensional model with some particular combination of contact and long-range interactions found recently \cite{2020BeauPRL}. In this context, broad research on effective numerical methods and techniques dedicated to different many-body systems are continuously ongoing\cite{1992LindenPhysRep,2000BeckPhysRep,2005SchollwockRMP,2007BartlettRMP,2008VerstraeteAdvPhys}.

Besides all sophisticated methods dedicated to quantum many-body systems, there is one strategy that is well-suited to the problem of finding isolated many-body ground states. The method is very natural since it is based on intuitive variational arguments \cite{2018GriffithsBook}. Simply, one needs to postulate a family of trail functions depending on a certain set of variational parameters. These parameters are optimized (analytically or numerically) to grant possibly minimal expectation value of the many-body Hamiltonian. If the trial family is luckily chosen appropriately, then the obtained wave function approximates the actual ground-state of the system very closely. Thus, all the ground-state properties are easily accessible with relatively low computation costs. Up to now, different approaches in this fashion have played a huge role in investigating different properties of various quantum systems.

It is worth emphasizing that direct insight into the mathematical structure of inter-particle correlations served by variational approaches should be recognized as a huge advantage having crucial importance. This is true especially if the state-of-the-art experiments with several ultra-cold atomic systems are considered \cite{2011SerwaneScience,2013WenzScience}. In this area, different experimental techniques enable one not only to precisely control and tune different parameters of a system but also to measure high-order correlations between particles with tremendous accuracy\cite{2020HoltenPRL}. Therefore, having well-justified theoretical predictions for the state's structure is a mandatory requirement for a clear understanding of the system properties\cite{2012BlumeRPP,2016ZinnerEPJ,2019SowinskiRPP}.

In the context of variational approaches applied to few-body problems, some effort was devoted recently to construct an appropriate variational trial wave function for particles confined in traps being close to parabolic shape\cite{2012RubeniPRA,2014WilsonPLA,2015LoftEPJD,2016BarfknechtJPhysB,2016AndersenSciRep,2017PecakPRA}. Starting from the well-known exact solution for two interacting particles\cite{2012BrouzosPRL}, following the Jastrow idea \cite{1995Jastrow}, there were proposed reasonable many-body wave functions for bosons\cite{2017KoscikFBS,2018KoscikEPL} as well as fermions\cite{2020KoscikNewJPhys}. Some sort of generalization to multi-component systems was also given\cite{2013BrouzosPRA,2018KoscikEPL}. All these attempts have one fundamental obstacle -- they cannot be straightforwardly adapted to other confinements since then they inappropriately capture the behavior of the system for strong interactions. In this work, by a deep investigation of a few-boson system confined in the external potential of an arbitrary shape, we show how to overcome this restriction. Notably, first, we elaborate on a two-particle problem and we define the ansatz that is exact in the limit of vanishing and infinite repulsions, while for intermediate interactions it nicely reproduces the actual ground-state of the system. Based on this experience, we show the strategy for obtaining the accurate ansatz for a larger number of particles. We demonstrate its effectiveness on different model systems with single- and double-well potentials over the entire interaction regime.

At this point, we want also to point out that the method presented here for confinements being far from the parabolic is not the only possible path to tackle the problem. For example, the problem of impurities confined in an asymmetric double-well potential was considered recently in the framework of the variational interpolatory ansatz\cite{2020LindgrenSciPost}.
 
\section{Formulation of the problem}\label{section1}
We consider a closed system of $N$ indistinguishable spinless bosons of mass $m$, interacting mutually via $\delta$-like contact interactions, and confined in a quasi-one-dimensional external trap. The Hamiltonian of the system has a form:
\begin{equation}\label{Hamiltonian_total}
 {\cal H} = \sum_{i=1}^N\left[-\frac{\hslash^2}{2m}\frac{\partial^2}{\partial x_i^2} + V(x_i)\right]+ g\sum_{i=1}^{N-1}\sum_{j=i+1}^N\delta(x_i-x_j),
\end{equation}
where $V(x)$ determines a shape of the external potential and $g$ controls the strength of interactions. Experimentally, such systems are attainable with ultra-cold dilute atomic gases \cite{2019SowinskiRPP}. A quasi-one-dimensional confinement is obtained by applying very strong confinement in two perpendicular directions and, besides utilization of the Feshbach resonances, may be used to control effective strength of interactions \cite{1998OlshaniiPRL}. 

In general, we aim to find the ground-state wave function of the system for possible general confinement $V(x)$. Although our strategy can be applied almost to any reasonable confinement, motivated by recent experimental implementations and theoretical proposals\cite{2006TheocharisPRE,2013HunnPRA,2013BugnionPRAb,2015MurmannPRLb,2016DobrzynieckiEPJD,2017CosmePRA,2018ErdmannPRA,2019ErdmannPRA}, in this work we focus on a quite general class of double-well potentials. They are modeled by a combination of a harmonic potential, a gaussian barrier in the middle, and a gradient field having a linear slope. It can be written as:
\begin{equation}\label{dw}
V(x)={m \omega^2 x^2\over 2} + \lambda \mathrm{e}^{-(x/\delta)^2/ 2}- \xi x + V_0.
\end{equation}
By controlling three independent parameters $\lambda$, $\delta$, and $\xi$ one can tune the trap to the pure harmonic oscillator ($\lambda=0$), symmetric ($\xi=0$) and non-symmetric ($\xi\neq 0$) double-well potential of controlled barrier height $\lambda$ and width $\delta>0$. To make further analysis clear, in the following we express all quantities in terms of natural units of the pure harmonic oscillator, {\it i.e.}, energy, length, and interaction strength are expressed in units of $\hslash \omega$, $\sqrt{\hslash/m\omega}$, and $\sqrt{\hslash^3 \omega/m}$, respectively. In this convention the parameters $\lambda$, $\delta$, and $\xi$ are dimensionless (they are expressed in units of energy, length, and force, respectively). For convenience, but without losing generality, we set the energy shift $V_0$ to the value ensuring the condition $V(0)=0$.

Before we present our variational scheme let us note here that even if the corresponding single-particle Hamiltonian is diagonalized and its eigenorbitals $\varphi_i(x)$ are known, the exact solution of this many-body problem is known only in a few specific cases. First, in the absence of interactions ($g=0$) all bosons occupy the lowest single-particle orbital $\varphi_0(x)$ and thus the ground-state wave function reads $\Psi(x_1,\ldots,x_N)=\prod_i\varphi_0(x_i)$. Second, for two bosons confined in a harmonic trap and arbitrary interactions the known solution has a form $\Psi(x_1,x_2)\sim \exp[-(x^2_++x_-^2)/2]U(\epsilon_g,1/2,x_-^2)$, where $x_\pm=(x_1\pm x_2)/\sqrt{2}$, $U(a,b,x)$ is the confluent hypergeometric function. The parameter $\epsilon_g$ is determined (via transcendental relation) by the interaction strength $g$. The detailed derivation of this solution can be found in the original paper by Busch {\it et al}\cite{1998BuschFoundPhys} or in a more pedagogical work \cite{2009WeiIJMPB}. Finally, as shown by Girardeau\cite{1960GirardeauJMP}, in the limit of infinitely strong repulsions ($g\rightarrow+\infty$), the $N$-boson ground-state wave function is also known exactly. Indeed, it can be directly obtained from the wave function of $N$ non-interacting fermions confined in the same external potential by applying an appropriate symmetrization. In all other cases, one can find the ground-state wave function only in an approximate fashion. For small or very strong interaction strengths it can be done perturbatively by expansion around known exact solutions or by performing numerical calculations based on the exact diagonalization. The most challenging (and therefore the most interesting) are of course cases with intermediate interactions where any expansion to any order becomes insufficient. In consequence, any numerical method is not reasonably well converged. 

\section{Variational ansatz}
Here we propose a relatively simple but in many cases very accurate variational ansatz for the bosonic wave function. It is based on the conviction that the dominant part of inter-particle correlations can be expressed by two-body correlations only. Since the bosonic wave function must be symmetric under the exchange of any two particles, this assumption leads directly to the most general form of the wave function:
\begin{equation}\label{Jastrow_ansatz}
 \Psi_G(x_1,\ldots,x_N) = {\cal N} \prod_{i=1}^{N}\left[\varphi(x_{i}) \prod_{j=i+1}^N\Phi(x_{i},x_{j})\right],
\end{equation} where $\varphi(x)$ and $\Phi(x,y)$ are some unknown single- and two-particle orbitals describing the state of the system and $\cal N$ is a numerical constant assuring appropriate normalization. This form of the many-body wave function can be viewed as a specific generalization of the celebrated Jastrow ansatz\cite{1995Jastrow} adapted to confined systems. Its specific form was used previously in the case of harmonic confinement for bosons as well as for fermions\cite{2018KoscikEPL,2020KoscikNewJPhys}. 

It can be shown straightforwardly\cite{1963LiebPR} that the contact interaction part of the Hamiltonian (\ref{Hamiltonian_total}) maybe directly included to the form of the wave function. Indeed, if the pair correlation function $\Phi(x,y)$ fulfills the condition
\begin{equation}\label{con1}
\left[\frac{\partial \Phi(x,y)}{\partial x}-\frac{\partial \Phi(x,y)}{\partial y}\right]\Bigr\vert_{x=y^{+}}-\left[\frac{\partial \Phi(x,y)}{\partial x}-\frac{\partial \Phi(x,y)}{\partial y}\right]\Bigr\vert_{x=y^{-}}=2g \Phi(x,y)\Bigr\vert_{x=y},
\end{equation}
then it is sufficient that the wave function $\Psi_G(x_1,\ldots,x_N)$ is found for the ground state of the non-interacting system, {\it i.e.}, the system described by the Hamiltonian
\begin{equation}
 {\cal H}_0 = \sum_{i=1}^N\left[-\frac{1}{2}\frac{\partial^2}{\partial x_i^2} + V(x_i)\right].
\end{equation}
Strictly speaking, if we build a variational family fulfilling the condition (\ref{con1}) then the variational approximation of the ground-state wave function is found by minimizing the non-interacting energy functional
\begin{equation}\label{functio}
E[\Psi_G(x_1,\ldots,x_N)]=\int \mathrm{d}x_1\cdots\mathrm{d}x_N\, \Psi^*_G(x_1,\ldots,x_N){\cal H}_0 \Psi_G(x_1,\ldots,x_N).
\end{equation} 

To propose the variational family as accurate as possible, first we rewrite the two-particle Jastrow orbital $\Phi(x,y)$ to a more convenient form. We express it in terms of yet unknown antisymmetric function $D(x,y)$ and a single real parameter $\alpha$ as
\begin{subequations} \label{PhiDef}
\begin{equation} \label{KoscikAnA}
\Phi_\alpha(x,y) = 1-\Lambda_\alpha(x,y) \mathrm{e}^{-\alpha \mathrm{sign}(x-y) D(x,y)},
\end{equation}
where 
\begin{equation}\label{pair}
\Lambda_\alpha(x,y)={g\over \alpha (\partial_x-\partial_y)D(x,y) +g}.
\end{equation}
\end{subequations}
By introducing above redefinition we automatically assure that the pair-correlation function $\Phi(x,y)$ fulfils contact condition (\ref{con1}) for arbitrary value of $\alpha$ and for arbitrary function $D(x,y)$ which is antisymmetric under exchange of variables, $D(x,y)=-D(y,x)$. In this way we included automatically all conditions forced by mutual interactions. Thus, we can use the family (after probing different functions $D(x,y)$) to minimise the energy functional (\ref{functio}). 

Although one has very wide freedom in choosing the antisymmetric function $D(x,y)$ and the single-particle orbital $\varphi(x)$, it is very convenient to tailor them in such a way that known exact results for particular systems are well-captured by the ansatz. Therefore, now we will impose conditions on the trial family forced by limiting cases. 

The simplest condition is imposed by considering the non-interacting limit ($g\rightarrow 0$). We know that in this limit all bosons occupy the lowest orbital $\varphi_0(x)$ of a corresponding single-particle problem. Thus, the variational wave function (\ref{Jastrow_ansatz}) captures this limit appropriately provided that the single-particle orbital $\varphi(x)$ is chosen as the lowest orbital $\varphi_0(x)$. Indeed, in the limit $g\rightarrow 0$ the variational parameter $\alpha\rightarrow +\infty$ and therefore the two-particle orbital becomes trivial, $\Phi(x,y)=1$.

Relatively more challenging is to fulfill conditions originating in the limit of infinite repulsions ($g\rightarrow+\infty$). It is clear that this limit is quite well mimicked by the ansatz for vanishing $\alpha$. Indeed, for $\alpha\rightarrow 0$ one finds $\Lambda_\alpha(x,y)\rightarrow 1$ and therefore (due to the expansion $\mathrm{e}^{-\alpha z}\approx 1 - \alpha z$) 
\begin{equation}\label{teur}
\Psi_{G}(x_1,\ldots,x_N)\approx {\cal N} \prod_{i=1}^{N}\left[\varphi_0(x_{i}) \prod_{j=i+1}^N \mathrm{sign}(x_i-x_j)D(x_{i},x_{j})\right].
\end{equation}
We know however that in this limit the exact form of the ground-state wave function can be obtained by exact mapping from the ground state of non-interacting fermions. It means that the exact bosonic many-body wave function is expressed as a product of the Slater determinant formed with the lowest single-particle orbitals and a chain of sign functions\cite{1960GirardeauJMP}:
\begin{equation}\label{TGexact}
\Psi_\mathrm{exact}(x_1,\ldots,x_N) \sim \prod_{i=1}^{N} \prod_{j=i+1}^N \mathrm{sign}(x_i-x_j)\mathrm{Det}[\varphi_{j}(x_{i})].
\end{equation}
Consequently, the pair-correlation function $D(x,y)$ should be chosen such that the function (\ref{teur}) may reconstruct the exact solution (\ref{TGexact}) as close as possible. Of course, there is no general prescription how to construct function $D(x,y)$ for arbitrary confinement. However, in the case of general double-well traps (\ref{dw}) studied here some rigorous hints can be formulated. The first hint clarifies when a pure harmonic trap potential is considered. Then the exact wave function has a simple form
\begin{equation}
\Psi_\mathrm{exact}^\mathrm{(HO)}(x_1,\ldots,x_N) \sim \prod_{i=1}^{N}\mathrm{e}^{-x_i^2/2} \prod_{j=i+1}^N |x_i-x_j|.
\end{equation}
Therefore, for harmonic confinement, by taking correlation function in a very simple form of $D(x,y)=x-y$ one reconstructs with (\ref{teur}) the exact solution {\it rigorously}. This kind of pair-correlation function was exploited in previous works\cite{2018KoscikEPL,2020KoscikNewJPhys}. The second hint can be found by considering the complementary system of two particles in the confinement of arbitrary shape. Then, the exact solution in the limit of infinite repulsions is expressed by two the lowest single-particle orbitals $\varphi_0(x)$ and $\varphi_1(x)$ as:
\begin{align}
\Psi_\mathrm{exact}^\mathrm{(2)}(x_1,x_2) &\sim |\varphi_0(x_1)\varphi_1(x_2)-\varphi_1(x_1)\varphi_0(x_2)| \nonumber \\
&=\varphi_{0}(x_{1})\varphi_{0}(x_{2})\left|\pi( x_{1})-\pi( x_{2})\right|,
\end{align}
where $\pi(x)=\varphi_1(x)/\varphi_0(x)$. It means that by choosing $D(x,y)=\pi(x)-\pi(y)$ in our variational scheme one reconstructs solution of the two-body problem in arbitrary trap for infinite repulsions {\it rigorously}. At this step it is also worth to consider further improvement for finite interactions which does not change applicability in the infinite repulsions limit. Namely, one can extend proposed correlation function by additional variational parameter $\beta$ responsible for an effective rescaling of single-particle orbitals and define \begin{equation} \label{Ansatz2P}
D_\beta(x,y) = \pi(\beta x)-\pi(\beta y).
\end{equation}
In this way, the ansatz is more flexible and may give improved results for intermediate interactions. It is worth noticing that in the case of pure harmonic confinement $\pi(x)=x$. It means that the parameter $\beta$ is redundant with rescaling $\alpha$ and therefore it becomes irrelevant. This all means that the proposed form of the correlation function (\ref{Ansatz2P}) defines very reasonable yet not very complicated ansatz for $N=2$ particles in almost any trap. In the following section, we discuss its advantages over the naive ansatz proposed previously based on a harmonic-like substitution $\pi(x)=x$.

In the case of a larger number of particles, one needs to take one step further in defining the variational ansatz and to release constraints for function $\pi(x)$ since it is no longer related to the ratio of the lowest two eigenfunctions of the single-particle problem. This step is essential since for any non-harmonic confinement and a large enough number of particles, the function $\pi(x)$ must gain substantial corrections from higher single-particle orbitals occupied by particles. Now we will show how to generalize the correlation function $D_\beta(x,y)$ to make it more flexible for different confinements but keeping its appropriateness in limiting cases discussed above.

Since in our work we consider only smooth and bounded potentials it is clear that function $\pi(x)$ can be expanded in Taylor series $\pi(x) = \sum_k p_k x^k$. It immediately means that, up to irrelevant constant factor $p_1 \beta$, the correlation function has a form
\begin{align} \label{expansionX}
D_\beta(x,y) &=\sum_{k=1}^\infty p_k\beta^k(x^k-y^k) \nonumber \\
& \sim (x-y) \left[1+\frac{p_2}{p_1}\beta(x+y)+\frac{p_3}{p_1}\beta^2(x^2+xy+y^2)+\ldots\right] =
 (x-y)\left[1+\sum_{k=2}^\infty \frac{p_{k}}{p_1}\beta^{k-1}{\cal P}_k(x,y)\right],
\end{align}
where ${\cal P}_k(x,y)$ is $(k-1)$-th order polynomial of the property $(x-y){\cal P}_k(x,y)=x^{k}-y^{k}$. Note that for any mirror-symmetric external potential ($\xi=0$) the function $\pi(x)$ is odd ($\pi(-x)=\pi(x)$). Then, only odd-$k$ terms are present in its decomposition and therefore the correlation function $D_\beta(x,y)$ is also simplified.

From the obtained structure of the correlation function originating in a two-orbital approach (\ref{expansionX}) we deduce possible paths of generalization. First, more accurate but numerically very challenging and numerical-time-consuming, relay on converting decomposition coefficients $p_k$ to independent variational parameters. In this approach, the final ansatz of $n$-th order is defined as
\begin{align} \label{expansionY}
D_{\vec{b}}(x,y) &\equiv (x-y) \left[1+\sum_{k=2}^{n+1}b_{k-1}{\cal P}_k(x,y)\right],
\end{align}
where $b_k$ is $k$-th element of variational parameters vector $\vec{b}$.

The second possibility, which we widely exploit in the following, goes in opposite direction. Namely, we associate all expansion coefficient $b_k$ with a single variational parameter $\beta$ (we set $b_k=\beta^k$) but we release a rigid polynomial structure of the expansion (\ref{expansionY}). Consequently we write the correlation function as 
\begin{align}
D_{\beta}(x,y) &\equiv (x-y) F_\beta(x,y),
\end{align}
where the variational function $F_\beta(x,y)$ obeys four conditions: {\it (i)} it is symmetric under exchange of variables, $F_\beta(x,y)=F_\beta(y,x)$; {\it (ii)} it saturates on unity for vanishing variational parameter $\beta$, $F_0(x,y)=1$; {\it (iii)} when Taylor-expanded in $\beta$, consecutive terms should be expressed as polynomials of $x$ and $y$ of consecutive powers; {\it (iv)} for mirror-symmetric potentials ($\xi=0$) only even polynomials contribute to the function $F_\beta(x,y)$, {\it i.e.}, terms with even powers of $\beta$ only. The conditions {\it (iii-iv)} are direct consequences of the explicit form of the expansion (\ref{expansionX}). These conditions mean that proper construction of the correlation function $F_\beta(x,y)$ should treat even and odd polynomial expressions as independent rather than related -- intensity of the odd part is controlled mostly by an asymmetry of external potential.

Now we are ready to propose the variational family of trial functions obeying desired conditions. In our work, we postulate one of the simplest forms which can be applied for any external potential having the form (\ref{dw}). It is constructed from two the lowest even polynomials and supplemented by the lowest odd polynomial to capture breaking of the mirror symmetry for cases with $\xi\neq 0$. The correlation function has a form
\begin{equation}\label{TheAnsatz}
D_\beta(x,y)=(x-y)\mathrm{e}^{-\beta^2(x^2+y^2)+\beta^4 (x^4+y^4)}\mathrm{e}^{-\gamma\beta(x+y)}.
\end{equation}
In this definition, mentioned independence of even and odd polynomials is provided by the additional parameter $\gamma$ which vanishes for mirror-symmetric scenarios. In general, it can be treated as an additional variational parameter. However, we checked that for a given number of particles and a given potential shape, it is sufficient to establish its value in the limiting case of infinite repulsions and then use it as the fixed parameter for all the other interactions. Of course, in the case of $N=2$ particles, it is not needed to use the ansatz (\ref{TheAnsatz}) since the previous approach based on the definition $D_\beta(x,y)=\pi(\beta x)-\pi(\beta y)$ gives similarly good results and is significantly much simpler in numerical adaptation. For more complicated confinements, like triple- or more-well potentials, one should try to extend the function $D_\beta(x,y)$ by higher-rank polynomial expressions in the exponent, keeping all the conditions for function $F_\beta(x,y)$ always fulfilled.

\section{Validation method}\label{section2}
To show that the proposed variational ansatz (\ref{TheAnsatz}) describes adequately properties of the many-body system confined in different traps we compare its predictions to the results obtained numerically by the exact diagonalization of the many-body Hamiltonian. For this purpose we use the optimised exact diagonalization method with the harmonic oscillator basis \cite{2018KoscikPhysLettA}. Since we deal with external confinements having parabolic shape far from the center, this approach assures that we get states having appropriate asymptotic behaviour for $x\rightarrow\pm \infty$. For a given number of particles $N$, interaction strength $g$, and parameters of the external confinement, the many-body Fock basis is constructed as follows. First we form a set of $K+1$ single-particle orbitals $\{\varphi^{\{\Omega\}}_0(x),\ldots,\varphi^{\{\Omega\}}_{K}(x)\}$ of the form 
\begin{equation}
\varphi^{\{\Omega\}}_k(x) =\frac{1}{\sqrt{2^k k!}}\left(\frac{\Omega}{\pi}\right)^{1/4} \mathrm{H}_k\left(\sqrt{\Omega} x\right) \mathrm{e}^{-\Omega x^2/2},
\end{equation}
where $\mathrm{H}_k(x)$ is the Hermite polynomial. Note that $\Omega$ plays a role of an optimalization parameter and is not dependent on frequency $\omega$ in (\ref{dw}). Then we construct the Fock basis over these states. Namely, in the second-quantized occupation notation the basis is formed by Fock states $|\mathtt{F}_\ell\rangle = |n_0,n_1,\ldots,n_K,0,\ldots\rangle$ (the index $\ell$ enumerates Fock states) with the cut-off condition $\sum_{k=0}^K k\,n_k < K$. In the next step, all matrix elements of the Hamiltonian $H_{\ell\ell'}=\langle \mathtt{F}_\ell|{\cal H}|\mathtt{F}_{\ell'}\rangle$ are calculated and the corresponding matrix is diagonalized. In this way approximate ground-state energy $E_0^{\{\Omega\}}$ and associated eigenvector representing the many-body ground state $|\mathtt{G}^{\{\Omega\}}\rangle$ are found. Finally, we repeat all these calculations for different values of the optimalization parameter $\Omega$ to find the lowest ground-state energy $E_0$ for a given cut-off parameter $K$. By increasing $K$, the final results are systematically improved and they converge to their ultimate (physical) values. 

In our comparison we focus on systems with $N\in\{2,3,4\}$ particles and the most natural quantity -- the single-particle reduced density matrix defined in the position representtion as
\begin{equation} \label{1RDM}
\rho(x,x')=\int \mathrm{d}x_2\cdots\mathrm{d}x_N\,\Psi^*_G(x,x_2,\ldots, x_N)\Psi_G(x',x_2,\ldots, x_N).
\end{equation}
This quantity encodes all possible results which can be obtained in arbitrary single-particle measurements. The diagonal part corresponds to the density profile of atomic cloud, while the off-diagonal elements of the matrix are responsible for spatial coherence of the system on a single-particle level. They can be quantified in many different ways. To deliver a reasonably complete comparison, in our work we consider three different cuts of the reduced density matrix: the diagonal, the anti-diagonal, and the horizontal. They are defined as
\begin{equation}
n(x) = \rho(x,x), \qquad m(x) = \rho(x,-x), \qquad h(x) = \rho(x,0).
\end{equation}
By comparing these three distributions obtained via the ansatz and the exact numerical calculations we can give a quite comprehensive validation test. 

\section{Two-particle case}
\begin{figure}[t]
\includegraphics[scale=0.15]{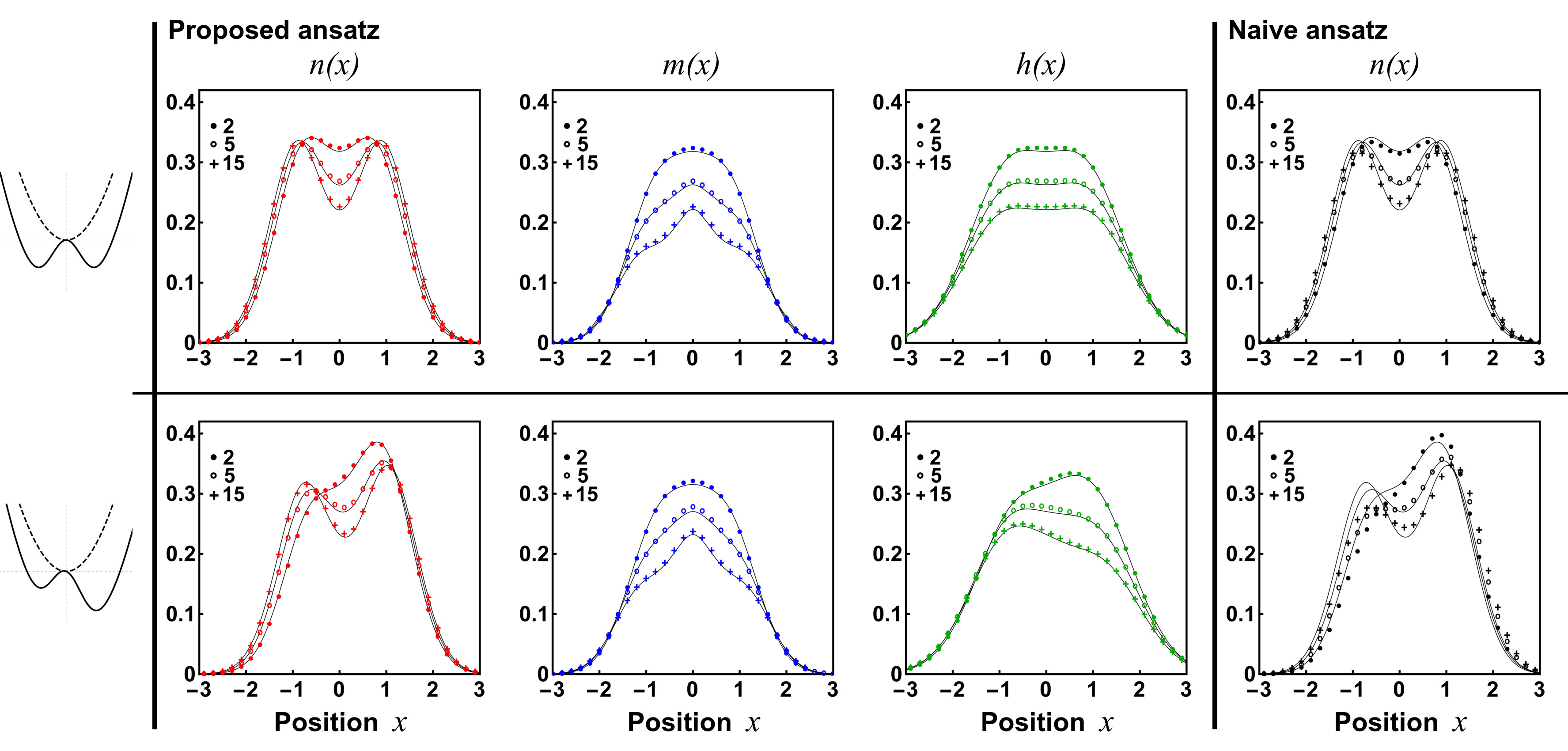}
\caption{Different cuts of the reduced single-particle density matrix for the system of $N=2$ bosons confined in different traps and with different interactions as predicted by the ansatz (different colour symbols) and compared with numerically exact results (solid lines). Consecutive columns present: the density profile $n(x)$, the anti-diagonal part $m(x)$, and the horizontal cut $h(x)$. Note almost perfect agreement between the ansatz's predictions and the exact results in a whole range of interactions as well as for traps of different symmetries. For comparison, in the last column we display the single-particle density profile $n(x)$ predicted by the naive ansatz with $D_\beta(x,y)=x-y$. In this case, significant deviations from the exact results (solid lines) are clearly visible. The two traps considered are characterized by $\lambda=1.0$, $\delta=0.5$ and the asymmetry $\xi=0.0$ and $\xi=0.2$, respectively. \label{Fig1}}
\end{figure}

We start validation of the ansatz in the simplest case of $N=2$ particles. As was mentioned previously, in this case it is not necessary to use the general ansatz (\ref{TheAnsatz}) but it is sufficient to utilize its simpler form $D_\beta(x,y) = \pi(\beta x)-\pi(\beta y)$ (see eq. (\ref{Ansatz2P})). In Fig.~\ref{Fig1}, we plot different cuts of the single-particle density matrix ($\rho(x)$, $m(x)$, and $h(x)$ on consecutive columns) predicted by the ansatz (different symbols) and we compare them with corresponding results obtained via the exact diagonalization (solid lines). To make a comparison comprehensive we analyze three different interaction strengths and two different double-well-like confinements (first and second row, respectively). It is clear that the ansatz correctly predicts shapes of all three distributions in a whole range of interactions not only in the case of symmetric double-well confinement ($\xi=0$) but also in the case of a very asymmetric scenario ($\xi=0.2$). For comparison, in the last column in Fig.~\ref{Fig1} we display predictions for the density distribution $\rho(x)$ of the same ansatz but with naive substitution $D_\beta(x,y) = x-y$ which was shown to be perfectly suited in the case of confinements close to the harmonic one\cite{2018KoscikEPL,2020KoscikNewJPhys}. It is rather obvious that the naive ansatz, even in the case of symmetric double-well confinement, is not optimal. Especially this is the case for positions being close to the double-well minima. The situation is even worse for asymmetric traps where this simplified ansatz is completely unreliable. This evident inappropriateness for asymmetric cases comes mainly from the fact that here the correlation function $D_\beta(x,y)$ shouldn't be mirror-symmetric. Mentioned breaking of mirror-symmetry is directly encoded in function $\pi(x)$ which is built from two of the lowest single-particle orbitals, while it is completely neglected by setting $D_\beta(x,y)=x-y$. 

Appropriateness of the ansatz studied and clear deviations of the naive approach can be visualized further when, instead of different cuts of the reduced single-particle density matrix, its global properties describing inter-particle correlations are considered. They are nicely quantified by the entanglement entropy $S$ which is one of natural measures of non-classical correlations between particles. Particularly, in the case of two indistinguishable particles, the general criterion of their entanglement based on the entanglement entropy was formulated \cite{2004GhirardiPRA}. It can be easily calculated if the decomposition of the reduced density matrix to its natural orbitals $\{u_i(x)\}$ is known. Namely, in the basis of natural orbitals, the density matrix $\rho(x,x')$ is expressed as a sum of projectors
\begin{equation}
\rho(x,x') = \sum_i r_i\, u^*_i(x)u_i(x'),
\end{equation}
where each $r_i$ is the eigenvalue of $\rho(x,x')$ associated with the eigenorbital $u_i(x)$. Then, having eigenvalues $\{r_i\}$ in hand, one defines the von Neumann entanglement entropy $S= -\sum_i r_{i}\,\mathrm{ln}\,r_{i}$. For the two-particle scenario considered here, we compute the eigenvalues $\{r_i\}$ directly from the wave function, {\it i.e.}, we diagonalize the matrix $\Psi_{ij}=[\Delta x\Psi_G(r_i,r_j)]$ on a dense grid $\Delta x= r_{i+1}-r_{i}$ and we find a set of its eigenvalues $\{k_i\}$. Then the eigenvalues of $\rho(x,x')$ are simply given as $ \lambda_ {i}=k_{i}^2$ (see for example \cite{2010KoscikPhysLettA}).  
\begin{figure}
\centering
\includegraphics[scale=0.15]{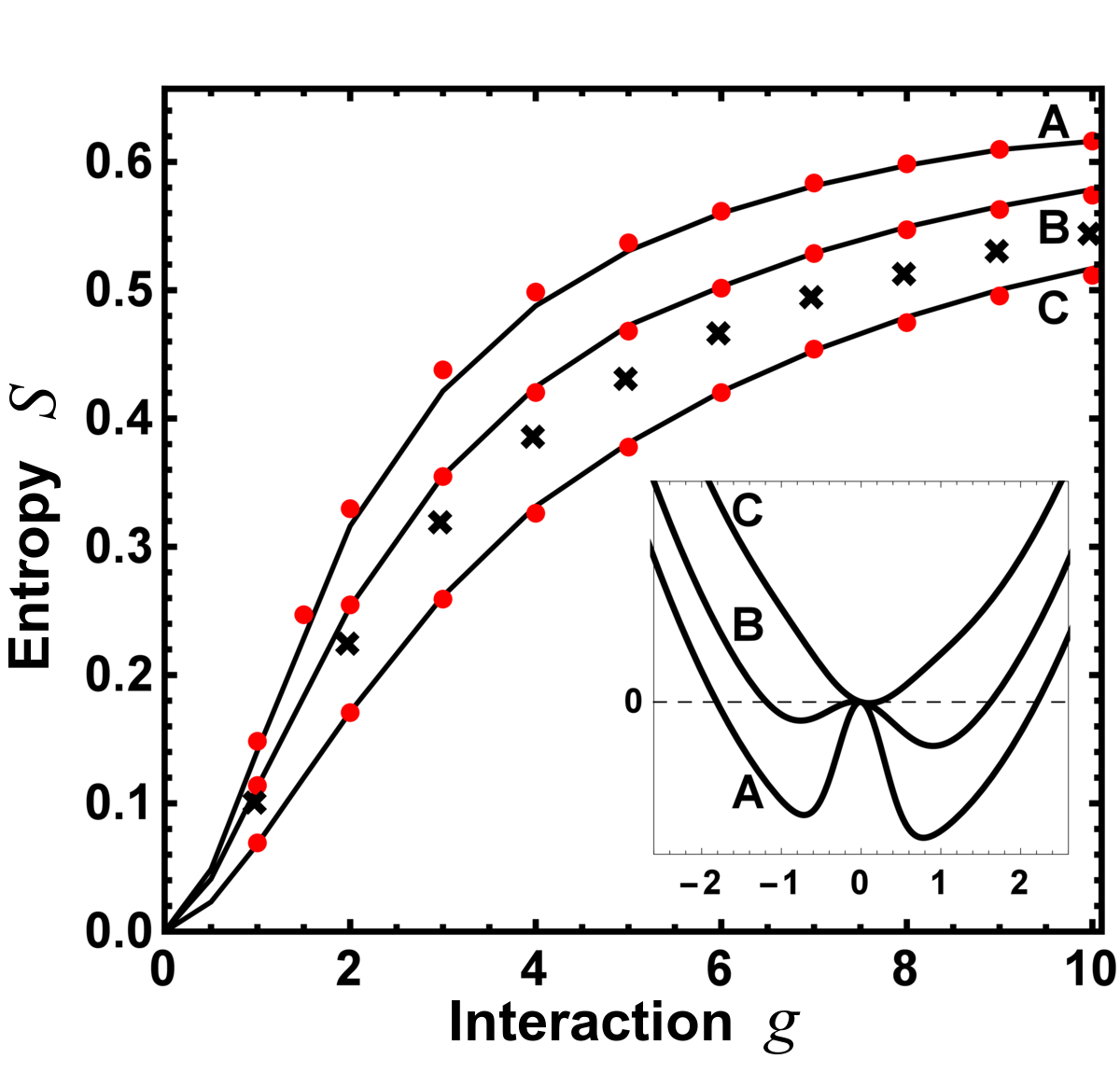}
\caption{Entanglement entropy $S$ calculated for the system with $N=2$ bosons encoded in the single-particle density matrix $\rho(x,x')$ as a function of interactions and for three different shapes of the confinement: $(\lambda,\delta,\xi)=(2.0,0.3,0.2)$, $(1.0,0.5,0.2)$, and $(-0.5,0.5,0.3)$ respectively for trap A, B, and C shown in inset. Exactly as in Fig.~\ref{Fig1}, predictions of the ansatz (red points) are consistent with numerically exact results (solid lines). Contrary, the naive ansatz (black crosses), based on substitution $D_\beta(x,y)=x-y$, is unreliable for strong enough repulsions. For clarity of the Figure we present results from the naive ansatz only for the asymmetric double-well potential labelled by B. For other confinements the results are analogous. \label{Fig2}}
\end{figure}

In Fig.~\ref{Fig2} we display the entanglement entropy $S$ as a function of mutual interactions $g$ for three different traps (displayed in the inset). We compare results obtained with two different approaches: the exact diagonalization (solid black lines) and the proposed variational ansatz (red symbols). Both approaches give almost the same results for a whole range of interactions and any external confinement considered. For comparison, in the case of shallow and asymmetric double-well confinement (trap {\bf B}) we also display results obtained with the naive approach based on substitution $D_\beta(x,y)=x-y$ (black crosses). In this case, deviations from the exact results and substantial underestimation of inter-particle correlations are clearly visible. This result supports previous findings based on the single-particle density profile. 

\section{Larger number of particles}
\begin{figure}[t]
\centering
\includegraphics[scale=0.15]{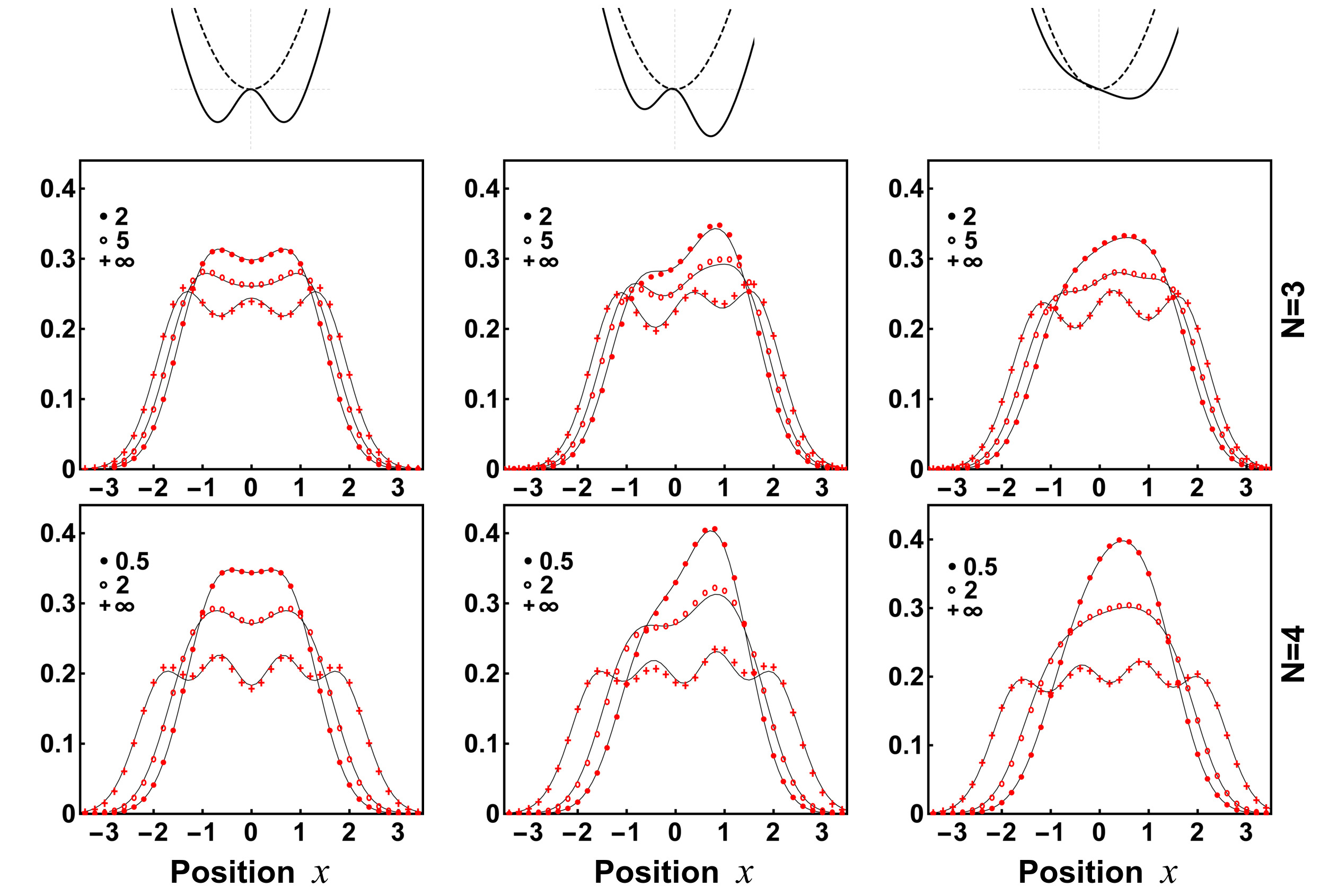}
\caption{Single-particle density profile $n(x)=\rho(x,x)$ for the system containing $N=3$ and $N=4$ bosons and different interaction strengths. Three consecutive columns correspond to three different types of the confinement with parameters $(\lambda,\delta,\xi)=(1.0,0.5,0.0)$, $(\lambda,\delta,\xi)=(1.0,0.5,0.2)$, and $(\lambda,\delta,\xi)=(1.0,1.0,0.2)$. Independently on the shape of the trap and interaction strength, the density profile predicted by the ansatz (red symbols) is in full agreement with numerically exact results (solid lines). \label{Fig3}}
\end{figure}
As argued before, in the case of a larger number of particles one needs to release stiffness of the ansatz and use a slightly more general family defined according to (\ref{TheAnsatz}). To show that the ansatz is very accurate also in these cases we perform a detailed analysis of systems containing $N=3$ and $N=4$ particles confined in three very different confinements being far from the parabolic trap: symmetric double-well potential, asymmetric double-well potential, and essentially asymmetric single-well trap (detailed parametrization is given in the caption of Fig.~\ref{Fig3}). In all these cases we investigate the accuracy of the ansatz for interactions for which still we are able to perform exact diagonalization of the Hamiltonian and thus establish an appropriate benchmark. Additionally, we also check predictions of the ansatz in the limit of infinite repulsions ($g\rightarrow\infty$) where the exact form of the many-body ground state is known. First, in Fig.~\ref{Fig3}, we display predictions for the diagonal part of the single-particle density matrix, {\it i.e.}, the density profile $n(x)$. It is clear that the ansatz's predictions are very close to the exact results for all three types of traps and any interaction strength considered. Note that along with increasing interactions the density distribution undergoes substantial transformations. This is especially the case for asymmetric traps where the density transits from a very asymmetric distribution (for weak interactions) to much more regular (for very strong interactions). Importantly, this behavior is perfectly captured by the ansatz. This very nice property of the variational scheme proposed is a direct consequence of its construction which by definition is tailored to reconstruct appropriately the limit of infinite repulsions. At this point let us also mention that obtaining numerically exact results via direct diagonalization of the Hamiltonian becomes numerically challenging when the number of particles is increased. Therefore, in the case $N=4$, we compare results for relatively weaker strengths than in the case of $N=3$ particles. Contrary, the numerical complexity of the ansatz approach is almost insensitive to the interaction strength and the number of particles. Thus, it can be successfully used for cases being far beyond the feasibility of the numerical diagonalization approach. 
 
\begin{figure}
\centering
\includegraphics[scale=0.15]{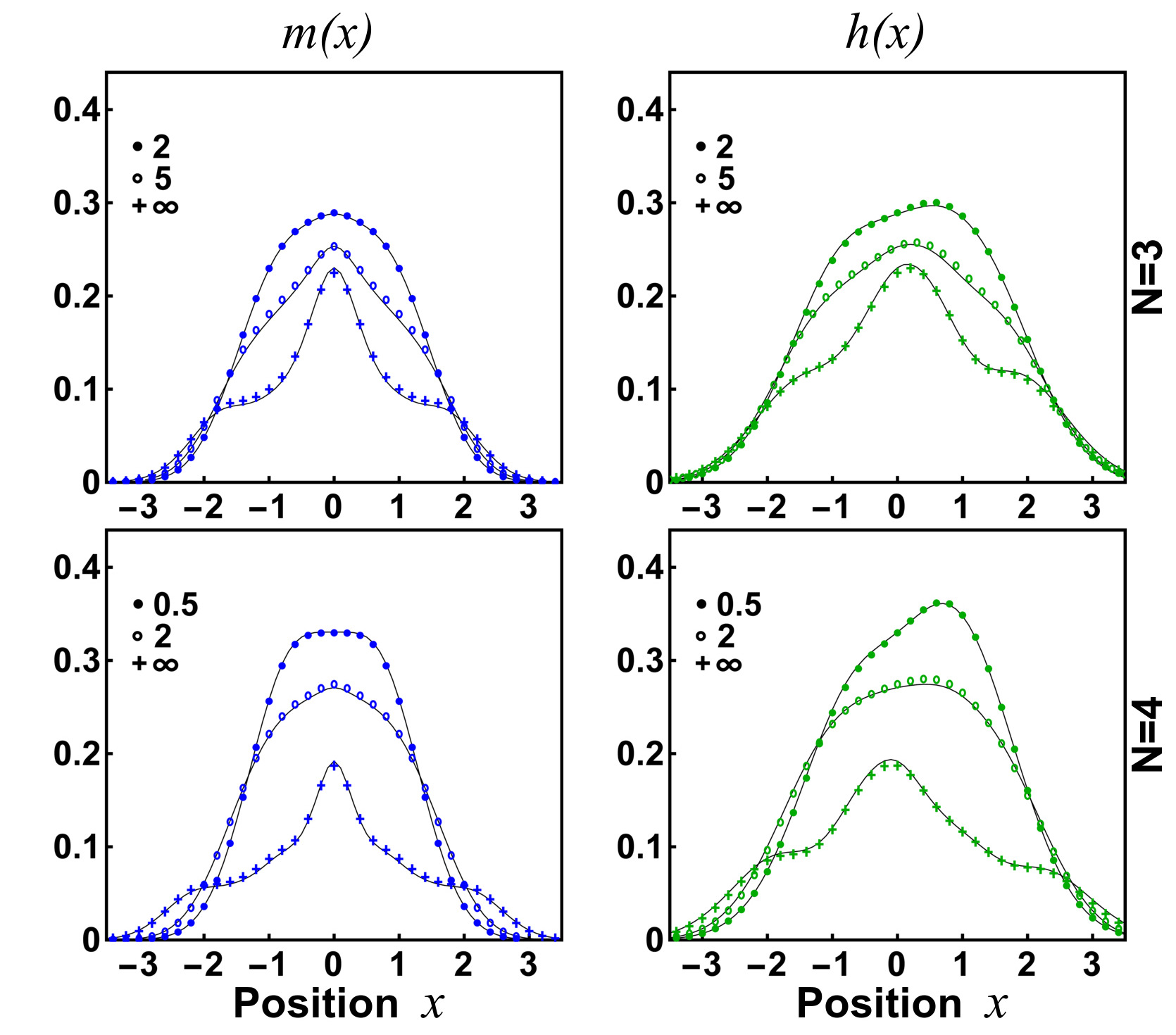}
\caption{Off-diagonal cuts of the reduced single-particle density matrix, $m(x)=\rho(x,-x)$ and $h(x)=\rho(x,0)$ for the system of $N=3$ and $N=4$ particles and different interaction strengths confined in an asymmetric double-well (corresponding to the second column in Fig.~\ref{Fig3}). Again, predictions of the ansatz are in full agreement with the numerically exact results. \label{Fig4}}
\end{figure}
Quality of the ansatz's predictions is also granted when instead of the diagonal part, different off-diagonal properties of the single-particle reduced density matrix are considered. We checked that two other quantities we focus on in our report, namely $m(x)$ and $h(x)$, are perfectly reconstructed by the ansatz for any shape of the external trap and any interaction strength. For example, in Fig.~\ref{Fig4}, we show corresponding results for the most challenging confinement -- asymmetric double-well trap (middle case in Fig.~\ref{Fig3}). For this trapping configuration, the exact diagonalization approach is on the border of feasibility and obtaining well-converged results requires huge numerical and computational resources. In contrast, the numerical complexity of the approach based on the ansatz (\ref{TheAnsatz}) does not substantially depend on the external trapping and even in this demanding case is easily manageable. 
 
At this point it is worth admitting that in contrast to other quantities, obtaining the von Neumann entanglement entropy $S$ from the ansatz is not an easy task. By the construction, the ansatz serves the ground-state as a wave function in the position domain. Thus, it is very easy to obtain all reduced quantities also in this representation. Entanglement entropy is not of this kind since it is obtained after diagonalization of the density matrix which in the position domain requires numerical integrations on a very dense grid. Thus, for systems with more than $N=2$ particles obtaining this particular quantity from the ansatz is not efficient.
 
\section{The ansatz as a convenient validator}
\begin{figure}
\centering
\includegraphics[scale=0.15]{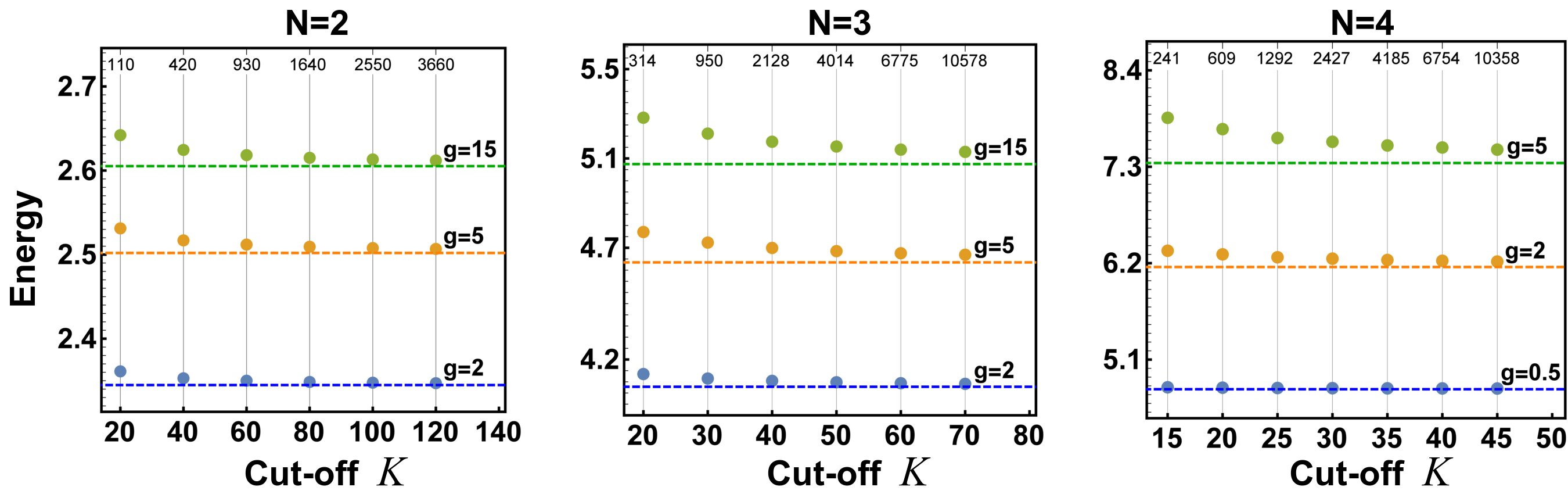}
\caption{Ground-state energy of the system of $N=2,3$, and $4$ particles as predicted by the ansatz (horizontal dashed lines) and as obtained by the numerically exact diagonalization of the Hamiltonian (\ref{Hamiltonian_total}) in the cropped Fock basis with cut-off $K$. Labels on thin vertical lines denote sizes of corresponding cropped Hilbert spaces. Note that along with increasing cut-off $K$ the numerically exact results converge from above to the ansatz's predictions. This indicates that the variational scheme proposed, although relatively simple and numerically less demanding, typically gives very accurate and better results than very heavy calculations in the framework of the matrix diagonalization. In the presented examples, calculations are performed for the same trap as in Fig.~\ref{Fig4} (corresponding to the second column in Fig.~\ref{Fig3}). \label{Fig5}}
\end{figure}
Up to now, our validation method was based on comparison with the results obtained via numerically exact diagonalization of the Hamiltonian. Of course, this method has limited capabilities since the diagonalization is always performed on a cropped basis formed by states with the lowest excitations. Thus it is not credible when highly-excited states have significant participation to the ground state. This is one of the fundamental reasons why in our previous comparisons we considered only weak enough interaction strengths for which the method is well-converged, {\it i.e.}, further inclusion of highly excited states does not change obtained results. To expose this issue, in Fig.~\ref{Fig5}, we compare the ground-state energy as predicted by the ansatz and the exact diagonalization method for different interactions and different numbers of particles. It is clear that along with an increasing number of single-particle orbitals $K$ from which the Fock basis is built, successive ground-state energies predicted by the exact diagonalization converge to their true values. At the same time, horizontal lines represent corresponding ground-state energies predicted by the ansatz. It is evident that typically the ansatz approximates the ground state much better than the diagonalization approach. Only for a small number of particles, relatively small interactions, and large enough Fock basis the ansatz estimates the ground-state energy slightly worse than the numerical diagonalization. It means that the ansatz proposed is not only much more convenient and faster than a direct diagonalization but in most cases gives results being closer to the physical reality. From this point of view, the ansatz should be recognized as an appropriate tool for the validation of other methods rather than the opposite.

\section{Conclusions}\label{section3}
In our work, we have shown how to get a very precise approximation of the many-body ground state of a one-dimensional system of several ultra-cold bosons in terms of the pair-correlated variational scheme. By detailed analysis of the contact interaction terms and a role played by external confinement, we give a straightforward recipe for appropriate construction of the variational family. Additionally, we have pointed out sources of potential discrepancies and we explained how to control them by including different variational terms. In principle, for a given external potential, the main idea is to treat the exact solution for infinite repulsions as a conductor benchmarking the accuracy of a whole method. The effectiveness of the approach proposed enabled us to conclude that the shape of the trapping potential strongly determines the spatial structure of particles' correlations, including the degree of entanglement. As consequence, we argued that other, recently used approaches \cite{2012BrouzosPRL,2018KoscikEPL,2020KoscikNewJPhys}, assuming pair-correlators as simple distance-dependent functions are insufficient for non-harmonic potentials. From this point of view, the presented results broaden the recent discussion on mutual correlations in systems of few ultra-cold atoms and open paths for additional searching for effective and accurate approximate methods. Although our work is devoted to the simplest problem of spinless bosons, the method presented can be generalized to the case of a larger number of components and also having different statistics, similarly as it was done previously for the naive ansatz with $D_\beta(x,y)=x-y$ \cite{2018KoscikEPL}. It can be done as long as the system considered is one-dimensional and mutual interactions have zero-range nature.

\section*{Acknowledgements }
This work was supported by the (Polish) National Science Centre Grant No. 2016/22/E/ST2/00555 (TS).

\section*{Author contributions statement}
P.K., A.K., A.P., and T.S. equally contributed in all stages of the project. All authors reviewed the manuscript.

\section*{Additional information}

\textbf{Competing financial and non-financial interests} All authors declare no competing interests.

\bibliography{biblio}

\begin{thebibliography}{10}
\expandafter\ifx\csname url\endcsname\relax
  \def\url#1{\texttt{#1}}\fi
\expandafter\ifx\csname urlprefix\endcsname\relax\def\urlprefix{URL }\fi
\providecommand{\bibinfo}[2]{#2}
\providecommand{\eprint}[2][]{\url{#2}}

\bibitem{1971FetterBook}
\bibinfo{author}{Fetter, A.~L.} \& \bibinfo{author}{Walecka, J.~D.}
\newblock \emph{\bibinfo{title}{Quantum Theory of Many-Particle Systems}}
  (\bibinfo{publisher}{McGraw-Hill}, \bibinfo{address}{Boston},
  \bibinfo{year}{1971}).

\bibitem{2008WeisseEDChapter}
\bibinfo{author}{Wei{\ss}e, A.} \& \bibinfo{author}{Fehske, H.}
\newblock \emph{\bibinfo{title}{Exact Diagonalization Techniques}},
  \bibinfo{pages}{529--544} (\bibinfo{publisher}{Springer Berlin Heidelberg},
  \bibinfo{address}{Berlin, Heidelberg}, \bibinfo{year}{2008}).
\newblock \urlprefix\url{https://doi.org/10.1007/978-3-540-74686-7_18}.

\bibitem{1998HaugsetPRA}
\bibinfo{author}{Haugset, T.} \& \bibinfo{author}{Haugerud, H.}
\newblock \bibinfo{title}{Exact diagonalization of the hamiltonian for trapped
  interacting bosons in lower dimensions}.
\newblock \emph{\bibinfo{journal}{Phys. Rev. A}} \textbf{\bibinfo{volume}{57}},
  \bibinfo{pages}{3809--3817} (\bibinfo{year}{1998}).
\newblock \urlprefix\url{https://link.aps.org/doi/10.1103/PhysRevA.57.3809}.

\bibitem{2007DeuretzbacherPRA}
\bibinfo{author}{Deuretzbacher, F.}, \bibinfo{author}{Bongs, K.},
  \bibinfo{author}{Sengstock, K.} \& \bibinfo{author}{Pfannkuche, D.}
\newblock \bibinfo{title}{Evolution from a bose-einstein condensate to a
  tonks-girardeau gas: An exact diagonalization study}.
\newblock \emph{\bibinfo{journal}{Phys. Rev. A}} \textbf{\bibinfo{volume}{75}},
  \bibinfo{pages}{013614} (\bibinfo{year}{2007}).
\newblock \urlprefix\url{https://link.aps.org/doi/10.1103/PhysRevA.75.013614}.

\bibitem{2018JeszenskiPRA}
\bibinfo{author}{Jeszenszki, P.}, \bibinfo{author}{Luo, H.},
  \bibinfo{author}{Alavi, A.} \& \bibinfo{author}{Brand, J.}
\newblock \bibinfo{title}{Accelerating the convergence of exact diagonalization
  with the transcorrelated method: Quantum gas in one dimension with contact
  interactions}.
\newblock \emph{\bibinfo{journal}{Phys. Rev. A}} \textbf{\bibinfo{volume}{98}},
  \bibinfo{pages}{053627} (\bibinfo{year}{2018}).
\newblock \urlprefix\url{https://link.aps.org/doi/10.1103/PhysRevA.98.053627}.

\bibitem{2018KoscikPhysLettA}
\bibinfo{author}{Ko{\'s}cik, P.}
\newblock \bibinfo{title}{Optimized configuration interaction approach for
  trapped multiparticle systems interacting via contact forces}.
\newblock \emph{\bibinfo{journal}{Physics Letters A}}
  \textbf{\bibinfo{volume}{382}}, \bibinfo{pages}{2561 -- 2564}
  (\bibinfo{year}{2018}).
\newblock
  \urlprefix\url{http://www.sciencedirect.com/science/article/pii/S0375960118306947}.

\bibitem{2019ChrostowskiAPPA}
\bibinfo{author}{Chrostowski, A.} \& \bibinfo{author}{Sowiński, T.}
\newblock \bibinfo{title}{Efficient construction of many-body fock states
  having the lowest energies}.
\newblock \emph{\bibinfo{journal}{Acta Physica Polonica A}}
  \textbf{\bibinfo{volume}{136}}, \bibinfo{pages}{566–570}
  (\bibinfo{year}{2019}).
\newblock \urlprefix\url{http://dx.doi.org/10.12693/APhysPolA.136.566}.

\bibitem{1968MoshinskyAJP}
\bibinfo{author}{Moshinsky, M.}
\newblock \bibinfo{title}{How good is the hartree-fock approximation}.
\newblock \emph{\bibinfo{journal}{American Journal of Physics}}
  \textbf{\bibinfo{volume}{36}}, \bibinfo{pages}{52--53}
  (\bibinfo{year}{1968}).
\newblock \urlprefix\url{https://doi.org/10.1119/1.1974410}.

\bibitem{1985BialynickiLetMPhys}
\bibinfo{author}{Bialynicki-Birula, I.}
\newblock \bibinfo{title}{Exact solutions of nonrelativistic classical and
  quantum field theory with harmonic forces}.
\newblock \emph{\bibinfo{journal}{Letters in Mathematical Physics}}
  \textbf{\bibinfo{volume}{10}}, \bibinfo{pages}{189--194}
  (\bibinfo{year}{1985}).
\newblock \urlprefix\url{https://doi.org/10.1007/BF00398157}.

\bibitem{1963LiebPR}
\bibinfo{author}{Lieb, E.~H.} \& \bibinfo{author}{Liniger, W.}
\newblock \bibinfo{title}{Exact analysis of an interacting bose gas. i. the
  general solution and the ground state}.
\newblock \emph{\bibinfo{journal}{Phys. Rev.}} \textbf{\bibinfo{volume}{130}},
  \bibinfo{pages}{1605--1616} (\bibinfo{year}{1963}).
\newblock \urlprefix\url{https://link.aps.org/doi/10.1103/PhysRev.130.1605}.

\bibitem{1963LiebPRb}
\bibinfo{author}{Lieb, E.~H.}
\newblock \bibinfo{title}{Exact analysis of an interacting bose gas. ii. the
  excitation spectrum}.
\newblock \emph{\bibinfo{journal}{Phys. Rev.}} \textbf{\bibinfo{volume}{130}},
  \bibinfo{pages}{1616--1624} (\bibinfo{year}{1963}).
\newblock \urlprefix\url{https://link.aps.org/doi/10.1103/PhysRev.130.1616}.

\bibitem{1966McGuireJMP1}
\bibinfo{author}{McGuire, J.~B.}
\newblock \bibinfo{title}{Interacting fermions in one dimension. i. repulsive
  potential}.
\newblock \emph{\bibinfo{journal}{Journal of Mathematical Physics}}
  \textbf{\bibinfo{volume}{6}}, \bibinfo{pages}{432--439}
  (\bibinfo{year}{1965}).
\newblock \urlprefix\url{https://doi.org/10.1063/1.1704291}.
\newblock \eprint{https://doi.org/10.1063/1.1704291}.

\bibitem{1966McGuireJMP2}
\bibinfo{author}{McGuire, J.~B.}
\newblock \bibinfo{title}{Interacting fermions in one dimension. ii. attractive
  potential}.
\newblock \emph{\bibinfo{journal}{Journal of Mathematical Physics}}
  \textbf{\bibinfo{volume}{7}}, \bibinfo{pages}{123--132}
  (\bibinfo{year}{1966}).
\newblock \urlprefix\url{https://doi.org/10.1063/1.1704798}.
\newblock \eprint{https://doi.org/10.1063/1.1704798}.

\bibitem{1967GaudinPLA}
\bibinfo{author}{Gaudin, M.}
\newblock \bibinfo{title}{Un systeme a une dimension de fermions en
  interaction}.
\newblock \emph{\bibinfo{journal}{Physics Letters A}}
  \textbf{\bibinfo{volume}{24}}, \bibinfo{pages}{55--56}
  (\bibinfo{year}{1967}).
\newblock
  \urlprefix\url{https://www.sciencedirect.com/science/article/pii/0375960167901934}.

\bibitem{1967YangPRL}
\bibinfo{author}{Yang, C.~N.}
\newblock \bibinfo{title}{Some exact results for the many-body problem in one
  dimension with repulsive delta-function interaction}.
\newblock \emph{\bibinfo{journal}{Phys. Rev. Lett.}}
  \textbf{\bibinfo{volume}{19}}, \bibinfo{pages}{1312--1315}
  (\bibinfo{year}{1967}).
\newblock \urlprefix\url{https://link.aps.org/doi/10.1103/PhysRevLett.19.1312}.

\bibitem{2020GamayunSciPost}
\bibinfo{author}{Gamayun, O.}, \bibinfo{author}{Lychkovskiy, O.} \&
  \bibinfo{author}{Zvonarev, M.~B.}
\newblock \bibinfo{title}{{Zero temperature momentum distribution of an
  impurity in a polaron state of one-dimensional Fermi and Tonks-Girardeau
  gases}}.
\newblock \emph{\bibinfo{journal}{SciPost Phys.}} \textbf{\bibinfo{volume}{8}},
  \bibinfo{pages}{53} (\bibinfo{year}{2020}).
\newblock \urlprefix\url{https://scipost.org/10.21468/SciPostPhys.8.4.053}.

\bibitem{1971CalogeroJMP}
\bibinfo{author}{Calogero, F.}
\newblock \bibinfo{title}{Solution of the one-dimensional n-body problems with
  quadratic and/or inversely quadratic pair potentials}.
\newblock \emph{\bibinfo{journal}{Journal of Mathematical Physics}}
  \textbf{\bibinfo{volume}{12}}, \bibinfo{pages}{419--436}
  (\bibinfo{year}{1971}).
\newblock \urlprefix\url{https://doi.org/10.1063/1.1665604}.

\bibitem{1971SutherlandJMP}
\bibinfo{author}{Sutherland, B.}
\newblock \bibinfo{title}{Quantum many-body problem in one dimension: Ground
  state}.
\newblock \emph{\bibinfo{journal}{Journal of Mathematical Physics}}
  \textbf{\bibinfo{volume}{12}}, \bibinfo{pages}{246--250}
  (\bibinfo{year}{1971}).
\newblock \urlprefix\url{https://doi.org/10.1063/1.1665584}.

\bibitem{2016BatchelorJPhysA}
\bibinfo{author}{Batchelor, M.~T.} \& \bibinfo{author}{Foerster, A.}
\newblock \bibinfo{title}{Yang{\textendash}baxter integrable models in
  experiments: from condensed matter to ultracold atoms}.
\newblock \emph{\bibinfo{journal}{Journal of Physics A: Mathematical and
  Theoretical}} \textbf{\bibinfo{volume}{49}}, \bibinfo{pages}{173001}
  (\bibinfo{year}{2016}).
\newblock \urlprefix\url{https://doi.org/10.1088/1751-8113/49/17/173001}.

\bibitem{2020BeauPRL}
\bibinfo{author}{Beau, M.}, \bibinfo{author}{Pittman, S.~M.},
  \bibinfo{author}{Astrakharchik, G.~E.} \& \bibinfo{author}{del Campo, A.}
\newblock \bibinfo{title}{Exactly solvable system of one-dimensional trapped
  bosons with short- and long-range interactions}.
\newblock \emph{\bibinfo{journal}{Phys. Rev. Lett.}}
  \textbf{\bibinfo{volume}{125}}, \bibinfo{pages}{220602}
  (\bibinfo{year}{2020}).
\newblock
  \urlprefix\url{https://link.aps.org/doi/10.1103/PhysRevLett.125.220602}.

\bibitem{1992LindenPhysRep}
\bibinfo{author}{{von der Linden}, W.}
\newblock \bibinfo{title}{A quantum monte carlo approach to many-body physics}.
\newblock \emph{\bibinfo{journal}{Physics Reports}}
  \textbf{\bibinfo{volume}{220}}, \bibinfo{pages}{53--162}
  (\bibinfo{year}{1992}).
\newblock
  \urlprefix\url{https://www.sciencedirect.com/science/article/pii/037015739290029Y}.

\bibitem{2000BeckPhysRep}
\bibinfo{author}{Beck, M.}, \bibinfo{author}{Jäckle, A.},
  \bibinfo{author}{Worth, G.} \& \bibinfo{author}{Meyer, H.-D.}
\newblock \bibinfo{title}{The multiconfiguration time-dependent hartree (mctdh)
  method: a highly efficient algorithm for propagating wavepackets}.
\newblock \emph{\bibinfo{journal}{Physics Reports}}
  \textbf{\bibinfo{volume}{324}}, \bibinfo{pages}{1--105}
  (\bibinfo{year}{2000}).
\newblock
  \urlprefix\url{https://www.sciencedirect.com/science/article/pii/S0370157399000472}.

\bibitem{2005SchollwockRMP}
\bibinfo{author}{Schollw\"ock, U.}
\newblock \bibinfo{title}{The density-matrix renormalization group}.
\newblock \emph{\bibinfo{journal}{Rev. Mod. Phys.}}
  \textbf{\bibinfo{volume}{77}}, \bibinfo{pages}{259--315}
  (\bibinfo{year}{2005}).
\newblock \urlprefix\url{https://link.aps.org/doi/10.1103/RevModPhys.77.259}.

\bibitem{2007BartlettRMP}
\bibinfo{author}{Bartlett, R.~J.} \& \bibinfo{author}{Musia\l{}, M.}
\newblock \bibinfo{title}{Coupled-cluster theory in quantum chemistry}.
\newblock \emph{\bibinfo{journal}{Rev. Mod. Phys.}}
  \textbf{\bibinfo{volume}{79}}, \bibinfo{pages}{291--352}
  (\bibinfo{year}{2007}).
\newblock \urlprefix\url{https://link.aps.org/doi/10.1103/RevModPhys.79.291}.

\bibitem{2008VerstraeteAdvPhys}
\bibinfo{author}{Verstraete, F.}, \bibinfo{author}{Murg, V.} \&
  \bibinfo{author}{Cirac, J.}
\newblock \bibinfo{title}{Matrix product states, projected entangled pair
  states, and variational renormalization group methods for quantum spin
  systems}.
\newblock \emph{\bibinfo{journal}{Advances in Physics}}
  \textbf{\bibinfo{volume}{57}}, \bibinfo{pages}{143--224}
  (\bibinfo{year}{2008}).
\newblock \urlprefix\url{https://doi.org/10.1080/14789940801912366}.

\bibitem{2018GriffithsBook}
\bibinfo{author}{Griffiths, D.~J.} \& \bibinfo{author}{Schroeter, D.~F.}
\newblock \emph{\bibinfo{title}{Introduction to quantum mechanics}}
  (\bibinfo{publisher}{Cambridge University Press}, \bibinfo{year}{2018}).

\bibitem{2011SerwaneScience}
\bibinfo{author}{Serwane, F.} \emph{et~al.}
\newblock \bibinfo{title}{Deterministic preparation of a tunable few-fermion
  system}.
\newblock \emph{\bibinfo{journal}{Science}} \textbf{\bibinfo{volume}{332}},
  \bibinfo{pages}{336--338} (\bibinfo{year}{2011}).
\newblock \urlprefix\url{http://science.sciencemag.org/content/332/6027/336}.

\bibitem{2013WenzScience}
\bibinfo{author}{Wenz, A.~N.} \emph{et~al.}
\newblock \bibinfo{title}{From few to many: Observing the formation of a fermi
  sea one atom at a time}.
\newblock \emph{\bibinfo{journal}{Science}} \textbf{\bibinfo{volume}{342}},
  \bibinfo{pages}{457--460} (\bibinfo{year}{2013}).
\newblock \urlprefix\url{http://science.sciencemag.org/content/342/6157/457}.

\bibitem{2020HoltenPRL}
\bibinfo{author}{Holten, M.} \emph{et~al.}
\newblock \bibinfo{title}{Observation of pauli crystals}.
\newblock \emph{\bibinfo{journal}{Phys. Rev. Lett.}}
  \textbf{\bibinfo{volume}{126}}, \bibinfo{pages}{020401}
  (\bibinfo{year}{2021}).
\newblock
  \urlprefix\url{https://link.aps.org/doi/10.1103/PhysRevLett.126.020401}.

\bibitem{2012BlumeRPP}
\bibinfo{author}{Blume, D.}
\newblock \bibinfo{title}{Few-body physics with ultracold atomic and molecular
  systems in traps}.
\newblock \emph{\bibinfo{journal}{Reports on Progress in Physics}}
  \textbf{\bibinfo{volume}{75}}, \bibinfo{pages}{046401}
  (\bibinfo{year}{2012}).
\newblock \urlprefix\url{https://doi.org/10.1088/0034-4885/75/4/046401}.

\bibitem{2016ZinnerEPJ}
\bibinfo{author}{{Zinner, Nikolaj Thomas}}.
\newblock \bibinfo{title}{Exploring the few- to many-body crossover using cold
  atoms in one dimension}.
\newblock \emph{\bibinfo{journal}{EPJ Web of Conferences}}
  \textbf{\bibinfo{volume}{113}}, \bibinfo{pages}{01002}
  (\bibinfo{year}{2016}).
\newblock \urlprefix\url{https://doi.org/10.1051/epjconf/201611301002}.

\bibitem{2019SowinskiRPP}
\bibinfo{author}{Sowi{\'{n}}ski, T.} \& \bibinfo{author}{Garc{\'{\i}}a-March,
  M.~{\'{A}}.}
\newblock \bibinfo{title}{One-dimensional mixtures of several ultracold atoms:
  a review}.
\newblock \emph{\bibinfo{journal}{Reports on Progress in Physics}}
  \textbf{\bibinfo{volume}{82}}, \bibinfo{pages}{104401}
  (\bibinfo{year}{2019}).
\newblock \urlprefix\url{https://doi.org/10.1088/1361-6633/ab3a80}.

\bibitem{2012RubeniPRA}
\bibinfo{author}{Rubeni, D.}, \bibinfo{author}{Foerster, A.} \&
  \bibinfo{author}{Roditi, I.}
\newblock \bibinfo{title}{Two interacting fermions in a one-dimensional
  harmonic trap: Matching the bethe ansatz and variational approaches}.
\newblock \emph{\bibinfo{journal}{Phys. Rev. A}} \textbf{\bibinfo{volume}{86}},
  \bibinfo{pages}{043619} (\bibinfo{year}{2012}).
\newblock \urlprefix\url{https://link.aps.org/doi/10.1103/PhysRevA.86.043619}.

\bibitem{2014WilsonPLA}
\bibinfo{author}{Wilson, B.}, \bibinfo{author}{Foerster, A.},
  \bibinfo{author}{Kuhn, C.}, \bibinfo{author}{Roditi, I.} \&
  \bibinfo{author}{Rubeni, D.}
\newblock \bibinfo{title}{A geometric wave function for a few interacting
  bosons in a harmonic trap}.
\newblock \emph{\bibinfo{journal}{Physics Letters A}}
  \textbf{\bibinfo{volume}{378}}, \bibinfo{pages}{1065--1070}
  (\bibinfo{year}{2014}).
\newblock
  \urlprefix\url{https://www.sciencedirect.com/science/article/pii/S0375960114001583}.

\bibitem{2015LoftEPJD}
\bibinfo{author}{{Loft, Niels Jacob S.}}, \bibinfo{author}{{Dehkharghani, Amin
  S.}}, \bibinfo{author}{{Mehta, Nirav P.}}, \bibinfo{author}{{Volosniev, Artem
  G.}} \& \bibinfo{author}{{Zinner, Nikolaj T.}}
\newblock \bibinfo{title}{A variational approach to repulsively interacting
  three-fermion systems in a one-dimensional harmonic trap}.
\newblock \emph{\bibinfo{journal}{Eur. Phys. J. D}}
  \textbf{\bibinfo{volume}{69}}, \bibinfo{pages}{65} (\bibinfo{year}{2015}).
\newblock \urlprefix\url{https://doi.org/10.1140/epjd/e2015-50845-9}.

\bibitem{2016BarfknechtJPhysB}
\bibinfo{author}{Barfknecht, R.~E.}, \bibinfo{author}{Dehkharghani, A.~S.},
  \bibinfo{author}{Foerster, A.} \& \bibinfo{author}{Zinner, N.~T.}
\newblock \bibinfo{title}{Correlation properties of a three-body bosonic
  mixture in a harmonic trap}.
\newblock \emph{\bibinfo{journal}{Journal of Physics B: Atomic, Molecular and
  Optical Physics}} \textbf{\bibinfo{volume}{49}}, \bibinfo{pages}{135301}
  (\bibinfo{year}{2016}).
\newblock \urlprefix\url{https://doi.org/10.1088/0953-4075/49/13/135301}.

\bibitem{2016AndersenSciRep}
\bibinfo{author}{Andersen, M. E.~S.}, \bibinfo{author}{Dehkharghani, A.~S.},
  \bibinfo{author}{Volosniev, A.~G.}, \bibinfo{author}{Lindgren, E.~J.} \&
  \bibinfo{author}{Zinner, N.~T.}
\newblock \bibinfo{title}{An interpolatory ansatz captures the physics of
  one-dimensional confined fermi systems}.
\newblock \emph{\bibinfo{journal}{Scientific Reports}}
  \textbf{\bibinfo{volume}{6}}, \bibinfo{pages}{28362} (\bibinfo{year}{2016}).
\newblock \urlprefix\url{https://doi.org/10.1038/srep28362}.

\bibitem{2017PecakPRA}
\bibinfo{author}{P{\c e}cak, D.}, \bibinfo{author}{Dehkharghani, A.~S.},
  \bibinfo{author}{Zinner, N.~T.} \& \bibinfo{author}{Sowi\'nski, T.}
\newblock \bibinfo{title}{Four fermions in a one-dimensional harmonic trap:
  Accuracy of a variational-ansatz approach}.
\newblock \emph{\bibinfo{journal}{Phys. Rev. A}} \textbf{\bibinfo{volume}{95}},
  \bibinfo{pages}{053632} (\bibinfo{year}{2017}).
\newblock \urlprefix\url{https://link.aps.org/doi/10.1103/PhysRevA.95.053632}.

\bibitem{2012BrouzosPRL}
\bibinfo{author}{Brouzos, I.} \& \bibinfo{author}{Schmelcher, P.}
\newblock \bibinfo{title}{Construction of analytical many-body wave functions
  for correlated bosons in a harmonic trap}.
\newblock \emph{\bibinfo{journal}{Phys. Rev. Lett.}}
  \textbf{\bibinfo{volume}{108}}, \bibinfo{pages}{045301}
  (\bibinfo{year}{2012}).
\newblock
  \urlprefix\url{https://link.aps.org/doi/10.1103/PhysRevLett.108.045301}.

\bibitem{1995Jastrow}
\bibinfo{author}{Jastrow, R.}
\newblock \bibinfo{title}{Many-body problem with strong forces}.
\newblock \emph{\bibinfo{journal}{Phys. Rev.}} \textbf{\bibinfo{volume}{98}},
  \bibinfo{pages}{1479--1484} (\bibinfo{year}{1955}).
\newblock \urlprefix\url{https://link.aps.org/doi/10.1103/PhysRev.98.1479}.

\bibitem{2017KoscikFBS}
\bibinfo{author}{Ko{\'{s}}cik, P.}
\newblock \bibinfo{title}{Fermionized dipolar bosons trapped in a harmonic
  trap}.
\newblock \emph{\bibinfo{journal}{Few-Body Systems}}
  \textbf{\bibinfo{volume}{58}}, \bibinfo{pages}{59} (\bibinfo{year}{2017}).
\newblock \urlprefix\url{https://doi.org/10.1007/s00601-017-1229-y}.

\bibitem{2018KoscikEPL}
\bibinfo{author}{Ko{\'{s}}cik, P.}, \bibinfo{author}{P{\l}odzie{\'{n}}, M.} \&
  \bibinfo{author}{Sowi{\'{n}}ski, T.}
\newblock \bibinfo{title}{Variational approach for interacting ultra-cold atoms
  in arbitrary one-dimensional confinement}.
\newblock \emph{\bibinfo{journal}{{EPL} (Europhysics Letters)}}
  \textbf{\bibinfo{volume}{123}}, \bibinfo{pages}{36001}
  (\bibinfo{year}{2018}).
\newblock \urlprefix\url{https://doi.org/10.1209%2F0295-5075%2F123%2F36001}.

\bibitem{2020KoscikNewJPhys}
\bibinfo{author}{Ko{\'{s}}cik, P.} \& \bibinfo{author}{Sowi{\'{n}}ski, T.}
\newblock \bibinfo{title}{Variational ansatz for p-wave fermions confined in a
  one-dimensional harmonic trap}.
\newblock \emph{\bibinfo{journal}{New Journal of Physics}}
  \textbf{\bibinfo{volume}{22}}, \bibinfo{pages}{093053}
  (\bibinfo{year}{2020}).
\newblock \urlprefix\url{https://doi.org/10.1088/1367-2630/abb386}.

\bibitem{2013BrouzosPRA}
\bibinfo{author}{Brouzos, I.} \& \bibinfo{author}{Schmelcher, P.}
\newblock \bibinfo{title}{Two-component few-fermion mixtures in a
  one-dimensional trap: Numerical versus analytical approach}.
\newblock \emph{\bibinfo{journal}{Phys. Rev. A}} \textbf{\bibinfo{volume}{87}},
  \bibinfo{pages}{023605} (\bibinfo{year}{2013}).
\newblock \urlprefix\url{https://link.aps.org/doi/10.1103/PhysRevA.87.023605}.

\bibitem{2020LindgrenSciPost}
\bibinfo{author}{Lindgren, E.~J.}, \bibinfo{author}{Barfknecht, R.~E.} \&
  \bibinfo{author}{Zinner, N.~T.}
\newblock \bibinfo{title}{{A systematic interpolatory method for an impurity in
  a one-dimensional fermionic background}}.
\newblock \emph{\bibinfo{journal}{SciPost Phys.}} \textbf{\bibinfo{volume}{9}},
  \bibinfo{pages}{5} (\bibinfo{year}{2020}).
\newblock \urlprefix\url{https://scipost.org/10.21468/SciPostPhys.9.1.005}.

\bibitem{1998OlshaniiPRL}
\bibinfo{author}{Olshanii, M.}
\newblock \bibinfo{title}{Atomic scattering in the presence of an external
  confinement and a gas of impenetrable bosons}.
\newblock \emph{\bibinfo{journal}{Phys. Rev. Lett.}}
  \textbf{\bibinfo{volume}{81}}, \bibinfo{pages}{938--941}
  (\bibinfo{year}{1998}).
\newblock \urlprefix\url{https://link.aps.org/doi/10.1103/PhysRevLett.81.938}.

\bibitem{2006TheocharisPRE}
\bibinfo{author}{Theocharis, G.}, \bibinfo{author}{Kevrekidis, P.~G.},
  \bibinfo{author}{Frantzeskakis, D.~J.} \& \bibinfo{author}{Schmelcher, P.}
\newblock \bibinfo{title}{Symmetry breaking in symmetric and asymmetric
  double-well potentials}.
\newblock \emph{\bibinfo{journal}{Phys. Rev. E}} \textbf{\bibinfo{volume}{74}},
  \bibinfo{pages}{056608} (\bibinfo{year}{2006}).
\newblock \urlprefix\url{https://link.aps.org/doi/10.1103/PhysRevE.74.056608}.

\bibitem{2013HunnPRA}
\bibinfo{author}{Hunn, S.}, \bibinfo{author}{Zimmermann, K.},
  \bibinfo{author}{Hiller, M.} \& \bibinfo{author}{Buchleitner, A.}
\newblock \bibinfo{title}{Tunneling decay of two interacting bosons in an
  asymmetric double-well potential: A spectral approach}.
\newblock \emph{\bibinfo{journal}{Phys. Rev. A}} \textbf{\bibinfo{volume}{87}},
  \bibinfo{pages}{043626} (\bibinfo{year}{2013}).
\newblock \urlprefix\url{https://link.aps.org/doi/10.1103/PhysRevA.87.043626}.

\bibitem{2013BugnionPRAb}
\bibinfo{author}{Bugnion, P.~O.} \& \bibinfo{author}{Conduit, G.~J.}
\newblock \bibinfo{title}{Exploring exchange mechanisms with a cold-atom gas}.
\newblock \emph{\bibinfo{journal}{Phys. Rev. A}} \textbf{\bibinfo{volume}{88}},
  \bibinfo{pages}{013601} (\bibinfo{year}{2013}).
\newblock \urlprefix\url{https://link.aps.org/doi/10.1103/PhysRevA.88.013601}.

\bibitem{2015MurmannPRLb}
\bibinfo{author}{Murmann, S.} \emph{et~al.}
\newblock \bibinfo{title}{Two fermions in a double well: Exploring a
  fundamental building block of the hubbard model}.
\newblock \emph{\bibinfo{journal}{Phys. Rev. Lett.}}
  \textbf{\bibinfo{volume}{114}}, \bibinfo{pages}{080402}
  (\bibinfo{year}{2015}).
\newblock
  \urlprefix\url{https://link.aps.org/doi/10.1103/PhysRevLett.114.080402}.

\bibitem{2016DobrzynieckiEPJD}
\bibinfo{author}{Dobrzyniecki, J.} \& \bibinfo{author}{Sowi{\'{n}}ski, T.}
\newblock \bibinfo{title}{Exact dynamics of two ultra-cold bosons confined in a
  one-dimensional double-well potential}.
\newblock \emph{\bibinfo{journal}{The European Physical Journal D}}
  \textbf{\bibinfo{volume}{70}}, \bibinfo{pages}{83} (\bibinfo{year}{2016}).
\newblock \urlprefix\url{https://doi.org/10.1140/epjd/e2016-70016-x}.

\bibitem{2017CosmePRA}
\bibinfo{author}{Cosme, J.~G.}, \bibinfo{author}{Andersen, M.~F.} \&
  \bibinfo{author}{Brand, J.}
\newblock \bibinfo{title}{Interaction blockade for bosons in an asymmetric
  double well}.
\newblock \emph{\bibinfo{journal}{Phys. Rev. A}} \textbf{\bibinfo{volume}{96}},
  \bibinfo{pages}{013616} (\bibinfo{year}{2017}).
\newblock \urlprefix\url{https://link.aps.org/doi/10.1103/PhysRevA.96.013616}.

\bibitem{2018ErdmannPRA}
\bibinfo{author}{Erdmann, J.}, \bibinfo{author}{Mistakidis, S.~I.} \&
  \bibinfo{author}{Schmelcher, P.}
\newblock \bibinfo{title}{Correlated tunneling dynamics of an ultracold
  fermi-fermi mixture confined in a double well}.
\newblock \emph{\bibinfo{journal}{Phys. Rev. A}} \textbf{\bibinfo{volume}{98}},
  \bibinfo{pages}{053614} (\bibinfo{year}{2018}).
\newblock \urlprefix\url{https://link.aps.org/doi/10.1103/PhysRevA.98.053614}.

\bibitem{2019ErdmannPRA}
\bibinfo{author}{Erdmann, J.}, \bibinfo{author}{Mistakidis, S.~I.} \&
  \bibinfo{author}{Schmelcher, P.}
\newblock \bibinfo{title}{Phase-separation dynamics induced by an interaction
  quench of a correlated fermi-fermi mixture in a double well}.
\newblock \emph{\bibinfo{journal}{Phys. Rev. A}} \textbf{\bibinfo{volume}{99}},
  \bibinfo{pages}{013605} (\bibinfo{year}{2019}).
\newblock \urlprefix\url{https://link.aps.org/doi/10.1103/PhysRevA.99.013605}.

\bibitem{1998BuschFoundPhys}
\bibinfo{author}{Busch, T.}, \bibinfo{author}{Englert, B.~G.},
  \bibinfo{author}{Rz\c{a}{\.z}ewski, K.} \& \bibinfo{author}{Wilkens, M.}
\newblock \bibinfo{title}{Two cold atoms in a harmonic trap}.
\newblock \emph{\bibinfo{journal}{Found. Phys.}} \textbf{\bibinfo{volume}{28}},
  \bibinfo{pages}{549} (\bibinfo{year}{1998}).

\bibitem{2009WeiIJMPB}
\bibinfo{author}{Wei, B.-B.}
\newblock \bibinfo{title}{Two one-dimensional interacting particles in a
  harmonic trap}.
\newblock \emph{\bibinfo{journal}{International Journal of Modern Physics B}}
  \textbf{\bibinfo{volume}{23}}, \bibinfo{pages}{3709--3715}
  (\bibinfo{year}{2009}).
\newblock \urlprefix\url{https://doi.org/10.1142/S0217979209053345}.

\bibitem{1960GirardeauJMP}
\bibinfo{author}{Girardeau, M.}
\newblock \bibinfo{title}{Relationship between systems of impenetrable bosons
  and fermions in one dimension}.
\newblock \emph{\bibinfo{journal}{Journal of Mathematical Physics}}
  \textbf{\bibinfo{volume}{1}}, \bibinfo{pages}{516--523}
  (\bibinfo{year}{1960}).
\newblock \urlprefix\url{https://doi.org/10.1063/1.1703687}.

\bibitem{2004GhirardiPRA}
\bibinfo{author}{Ghirardi, G.} \& \bibinfo{author}{Marinatto, L.}
\newblock \bibinfo{title}{General criterion for the entanglement of two
  indistinguishable particles}.
\newblock \emph{\bibinfo{journal}{Phys. Rev. A}} \textbf{\bibinfo{volume}{70}},
  \bibinfo{pages}{012109} (\bibinfo{year}{2004}).
\newblock \urlprefix\url{https://link.aps.org/doi/10.1103/PhysRevA.70.012109}.

\bibitem{2010KoscikPhysLettA}
\bibinfo{author}{Kościk, P.} \& \bibinfo{author}{Okopińska, A.}
\newblock \bibinfo{title}{Two-electron entanglement in elliptically deformed
  quantum dots}.
\newblock \emph{\bibinfo{journal}{Physics Letters A}}
  \textbf{\bibinfo{volume}{374}}, \bibinfo{pages}{3841--3846}
  (\bibinfo{year}{2010}).
\newblock
  \urlprefix\url{https://www.sciencedirect.com/science/article/pii/S0375960110009199}.

\end{thebibliography}

\end{document}